\documentclass[onecolumn,secnumarabic,amssymb,superscriptaddress,nobibnotes,aps,prd]{revtex4-1}

\usepackage[colorlinks, citecolor=blue]{hyperref}
\usepackage[fleqn]{amsmath}
\usepackage{graphics}
\usepackage{xcolor}
\usepackage{bm}
\usepackage{extarrows}
\usepackage{mathrsfs}
\usepackage{empheq}

\def\rar{\rightarrow}

\def\ra{\rangle}
\def\la{\langle}

\def\no{\nonumber}
\def\bea{\begin{eqnarray}}
\def\eea{\end{eqnarray}}
\def\be{\begin{equation}}
\def\ee{\end{equation}}

\def\p{\partial}

\begin{document}

\title{Quantum estimation in an expanding spacetime}%

\author{Xiaoyang Huang}
\affiliation{School of Science, Xi'an Jiaotong University, Xi'an, Shaanxi 710049, China}

\author{Jun Feng}%
\email{j.feng@xjtu.edu.cn}
\affiliation{School of Science, Xi'an Jiaotong University, Xi'an, Shaanxi 710049, China}
\affiliation{School of Mathematics and Physics, The University of Queensland, Brisbane, QLD 4072, Australia}

\author{Yao-Zhong Zhang}%
\affiliation{School of Mathematics and Physics, The University of Queensland, Brisbane, QLD 4072, Australia}
\affiliation{Institute of Modern Physics, Northwest University, Xi'an, Shaanxi 710069, China}

\author{Heng Fan}%
\affiliation{Beijing National Laboratory for Condensed Matter Physics, Institute of Physics, Chinese Academy of Sciences, Beijing 100190, China}
\date{\today}%

\begin{abstract}

We investigate the quantum estimation on the Hubble parameter of an expanding de Sitter space by quantum metrological techniques. By exploring the dynamics of a freely falling Unruh-DeWitt detector, which interacts with a scalar field coupling to curvature, we calculate the Fisher information (FI) and quantum Fisher information (QFI) for the detector, which bound the highest precision of the estimation on Hubble parameter. In standard Bunch-Davies vacuum, we show that the maxima of FI/QFI are located for particular initial state of probe. Beside its dependence on the evolving time of detector and the energy spacing of atom $\omega$, we show that the maxima of FI/QFI can be significantly enhanced once a proper coupling of scalar field to curvature is chosen. For instance, we show numerically that the estimation in the scenario with minimally/nearly minimally coupling scalar field can always outperform that with conformally coupling scalar field, corresponding to a higher FI/QFI in estimation. Moreover, we find that for general $\alpha-$vacua of de Sitter space, a further improvement of estimation can be achieved, attributed to the squeezed nature of $\alpha-$vacua that heavily constrains the measurement uncertainty. Some implications of our results are also discussed.

\end{abstract}


\maketitle

\section{Introduction}
\label{1}

Quantum metrology, using nonclassical properties of the probes, promises an estimation of physical parameters with high-precision superior to any classical procedure \cite{Met2}. The techniques of quantum parameter estimation have been developed in several metrology platforms, such as optical interferometry \cite{Met3}, cold atomic systems \cite{Met6}, and Bose-Einstein condensates \cite{Met8}. At large scale of spacetime where relativity is relevant, quantum-enhanced metrology is also applied to probe various extremely sensitive relativistic effects. The field began with the proposal of detecting gravitational wave (GW) with quantum measurement strategy decades ago \cite{Met9}, which arrives its thriving by LIGO's directly observation of GW in 2015 \cite{Met10}. 
Other applications of quantum metrology in relativity include witnessing the time dilation in quantum clock synchronization \cite{Met12,Met13}, and atom interferometry in a micro-gravity environment \cite{Met0}. Nevertheless, it is worth mentioning that most of these schemes are constructed from non-relativistic quantum mechanics which is incompatible with relativity.

In general, a parameter $\theta$ can be estimated with high-accuracy when the states $\rho_\theta$ and $\rho_{\theta+d\theta}$ can be distinguished for infinitesimal change $d\theta$. The operational measure of the distinguishability is the classical Fisher information (FI) $\mathcal{F}_C(\theta)$ \cite{FI1}, which essentially quantifies the amount of information that can be extracted about the unknown parameter $\theta$. In fact, FI gives us a lower bound to the mean-square error in the estimation through Cram\'er-Rao inequality \cite{FI2} $\mbox{Var}(\theta)\geqslant [N\mathcal{F}_C(\theta)]^{-1}$, where $N$ is the number of repeated experiments. Optimizing over all possible quantum measurements, an even stronger lower bound as $\mbox{Var}(\theta)\geqslant [N\mathcal{F}_Q(\theta)]^{-1}$ can be provided \cite{FI3}, where $\mathcal{F}_Q(\theta)$ is the quantum Fisher information (QFI) satisfying $\mathcal{F}_Q(\theta) \geqslant \mathcal{F}_C(\theta)$. Therefore, to attain the highest-precision in quantum metrological scheme, how to increase the QFI and preserve it against environment noise are key issues to be explored \cite{FI4,FI5}.

In recent years, there has been a growing ambition in applying quantum metrology techniques to probe quantum gravitational effects. A prominent example is Unruh effect \cite{Unruh1}, which claims that a uniformly accelerated detector interacting with external fields becomes excited in Minkowski vacuum. Essentially, this can be attributed to the fact that the concept of particle is observer-dependent for relativistic quantum fields \cite{QFT}. This profound insight has also lead to the celebrated Hawking radiation which indicates the thermal nature of black hole \cite{Unruh2}. Therefore, in order to exploit optimal metrology strategy for quantum gravity, it is natural to incorporate relativistic quantum fields to quantum metrology \cite{Met19,Met16,Met17}, thus leads to so-called relativistic quantum metrology (RQM). While the temperature of Unruh radiation is extremely weak (smaller than 1K for accelerations up to $10^{21}$$m/s^2$), it was shown \cite{Met16,Met15} that the relativistic feature of quantum fields, localized in moving cavity, can improve the QFI for measurement of Unruh temperature, at acceleration within reach of current experiments. Thus it gets an optimal precision higher than its non-relativistic counterpart. By employing Unruh-DeWitt detector model, it was shown \cite{Met20,Met18} that once multipartite entanglement exists, the metrological accuracy for parameter estimation of quantum gravity can be further improved to beat quantum limit. For general spacetime, the estimation on the spacetime geometry via relativistic metrological tasks have also been discussed \cite{Met21,Met22,Met23}. Moreover, with great ambition, applications of quantum measurement on probing new physics at Planck-scale, e.g., the generalized uncertainty relation, have been proposed in Refs. \cite{Met24,Met25}.

In this paper, we explore the relativistic quantum metrological tasks in de Sitter space, which forms the basis of inflationary cosmology and approximates the current accelerating universe \cite{ds1}. We employ \emph{local} quantum estimation theory to estimate the Hubble parameter $H$ of de Sitter space, which sets a positive cosmological constant $\Lambda=3H^2$ and the curvature $R=12H^2$. Moreover, resembling to Unruh effect in flat spacetime, in de Sitter space, any freely falling observer can perceive thermal radiation of Gibbons-Hawking temperature, related to Hubble parameter through $T_H=H/2\pi$ \cite{Unruh3}. Therefore, by an optimal quantum measurement on the Hubble parameter with high-precision, abundant information on the geometry and dynamics of the expanding de Sitter space can be extracted. 

In our detection model, the probe of estimation is modeled by an inertial Unruh-DeWitt detector, which couples to curved background through a scalar field and behaves like an open quantum system, while the fluctuations of the quantum field is treated as the environment \cite{Met17,Met20}. As is well known, relativistic quantum fields in de Sitter space exhibit more subtle behavior than they live in flat spacetime. For instance, to determine the dynamics of detector, besides its energy level spacing and the field mass, the couplings between scalar field and curvature should also be specified \cite{ds2,ds3}. On the other hand, quantum fields in de Sitter space are further complicated by the existence of abounding vacuum states (so-called $\alpha-$vacua) which all respect the symmetries of de Sitter space \cite{ds4,ds5}. Among them, the unique Bunch-Davies vacuum (as $\alpha\rar-\infty$)  \cite{ds6} is often used in cosmology as an initial state of inflationary era, while the rest general $\alpha-$vacua are intimately connected to trans-Planckian physics on the CMBR anisotropies produced by inflation \cite{ds7,ds8} and to de Sitter holography \cite{ds9,ds10}. Moreover, the profound entanglement exhibited in general $\alpha-$vacua, e.g., the entanglement entropy of quantum fields \cite{ds12,ds13} and its influence on various quantum information tasks \cite{ds14,ds15,dsplus}, has been intensively studied. 

In this paper, we give an analytically calculation of the QFI of quantum estimation on Hubble parameter. We find that the precision of estimation is very sensitive to the choice of initial state of the probe, the energy gap of the detector, as well as the couplings between the scalar field and de Sitter background. We also investigate the QFI in terms of choice of general de Sitter vacuum states. Since an $\alpha$-vacuum can be interpreted as a squeezed state over Bunch-Davies vacuum \cite{ds11}, resembling the case of quantum optics, a heavily constrain on quantum uncertainty can be expected. We show that such a general choice of $\alpha$ can give a significant enhancement on QFI, which leads to higher precision of quantum estimation than Bunch-Davies choice. In inflationary paradigm, specific $\alpha-$vacuum should be imposed as initial condition of inflaton fluctuation, that manifest possible short-distance cutoff at some fundamental scales of new physics. In such scenario, it is then impossible to prepare the initial $\alpha-$vacua in order to improve quantum measurements. Conversely, one can employ the optimal QFI as an indicator to distinguish different initial conditions imposed.  

The paper is organized as follows. In section \ref{2}, we introduce our detection model with an Unruh-DeWitt detector, modeled by a two-level atom, interacts with a free scalar field coupling to de Sitter background. By resolving the master equation of Lindblad form, we obtain the density matrix relating to the final state of detector. In section \ref{3}, we investigate RQM for various couplings $\xi$ between curvature and scalar fields, corresponding to different dynamics. In particular, we calculate and compare the FI and QFI for massless scalar field with conformally coupling ($\xi=1/6,m_\phi=0$) and minimally coupling ($\xi=0,m_\phi=0$), and nearly minimally coupled light scalar ($|\xi|\ll1,m_\phi\ll H$), respectively. In section \ref{4}, we explore that how a general choice of $\alpha-$vacua for quantum dynamics of Unruh-DeWitt detector can enhance the precision of quantum metrological tasks.  
Finally, a summary of our results is given in section \ref{6}. Throughout the paper, we use the natural units as $\hbar=c=G=k_B=1$.

\section{Detection model}
\label{2}

To proceed, we model an Unruh-DeWitt detector by a two-level atom which interacts a bath of fluctuating quantum scalar field in de Sitter space. This indicates that the detector behaves like an open system, while the fluctuations of the quantum field is treated as the environment \cite{op1,op2}. The full dynamics of detector then can be obtained by tracing over all field degrees of freedom, which effectively lead to environment decoherence and dissipation on the quantum state of system.

Without loss of generality, the total Hamiltonian of the combined system is 
\be
H=H_{s}+H_\Phi+\mu H_I\label{model}
\ee
where $H_s=\frac{1}{2}\omega \sigma_3$ and $H_\Phi$ are the Hamiltonian of detector and free scalar field $\Phi(x)$ in de Sitter space, respectively. The detector-field interaction is described by $H_I=(\sigma_++\sigma_-)\Phi(x)$.  Here, we denote the atomic raising/lowering operators as $\sigma_\pm$, and $\omega$ is the energy level spacing of the atom. 

The time evolution of the total system in the proper time $\tau$ of the detector is governed by von Neumann equation $\p_\tau \rho_{tot}(\tau)=-i[H,\rho_{tot}(\tau)]$, where the initial state can be taken as $\rho_{tot}(0)=\rho(0)\otimes |0\ra\la0|$, as $\rho(0)$ is the initial state of detector and $|0\ra$ is a vacuum state of scalar field respecting de Sitter symmetry. Following the approach of \cite{op1}, we assume a weak coupling limit ($\mu\ll1$). Then the reduced dynamics of the density matrix of detector $\rho(\tau)$ should evolve in time according to a one-parameter quantum dynamical semigroup of completely positive map, generated by the Kossakowski-Lindblad form
\be
\frac{\p \rho(\tau)}{\p \tau}=-i[H_{\mbox{\tiny eff}},\rho(\tau)]+L[\rho(\tau)]\label{ds1}
\ee
where
\be
L[\rho]=\frac{1}{2}\sum^3_{i,j=1}C_{ij}[2\sigma_j\rho \sigma_i-\sigma_i\sigma_j\rho-\rho\sigma_i\sigma_j]
\ee
is a nonunitary evolution term produced by the coupling with the external fields. The Kossakowski matrix $C_{ij}$ can be determined by Fourier transform $\mathcal{G}(\lambda)$ of the Wightman functions of scalar field $G^+(x-y)=\la0|\Phi(x)\Phi(y)|0\ra$, which gives
\be
\mathcal{G}(\lambda)=\int_{-\infty}^{\infty}d\tau~e^{i\lambda\tau}G^+(x(\tau))\label{ds8}
\ee
In particular, coefficients $C_{ij}$ can be written explicitly as \cite{op3,op4}
\be
C_{ij}=A\delta_{ij}-iB\epsilon_{ijk} n_k+Cn_in_j
\ee
where $n_i$ are the components of a unit vector, and
\be
A=\frac{1}{2}[\mathcal{G}(\omega)+\mathcal{G}(-\omega)],\quad B=\frac{1}{2}[\mathcal{G}(\omega)-\mathcal{G}(-\omega)],\quad C=\mathcal{G}(0)-A
\label{ds11}
\ee
Moreover, the interaction with external scalar field would also induce a Lamb shift contribution for the detector effective Hamiltonian $H_{\mbox{\tiny eff}}=\frac{\Omega}{2}\sigma_3$, in terms of a renormalized frequency $\Omega=\omega+i[\mathcal{K}(-\omega)-\mathcal{K}(\omega)]$, where $\mathcal{K}(\omega)$ is Hilbert transform of Wightman functions defined by $\mathcal{K}(\lambda)=\frac{1}{i\pi}\mbox{P}\int_{-\infty}^{\infty}d\omega\frac{\mathcal{G}(\omega)}{\omega-\lambda}$. 

To resolve equation (\ref{ds1}), we express the density matrix of detector in terms of Pauli matrices
\be
\rho(\tau)=\frac{1}{2}\Big(1+\sum_{i=1}^3\rho_i(\tau)\sigma_i\Big)\label{ds2}
\ee 
With a general initial state of detector as $|\psi\ra=\sin\frac{\theta}{2}|0\ra+\cos\frac{\theta}{2}|1\ra$, the time-dependent components of Bloch vector $\bm{\rho}\equiv\{\rho_1(\tau),\rho_2(\tau),\rho_3(\tau)\}$ can be analytically given as
\bea
\rho_1&=&e^{-\frac{1}{2}A\tau}\sin\theta\cos\Omega\tau\no\\
\rho_2&=&e^{-\frac{1}{2}A\tau}\sin\theta\sin\Omega\tau\no\\
\rho_3&=&e^{-A\tau}\cos\theta-R(1-e^{-A\tau})\label{ds3}
\eea
with $R$ is the ratio $R=B/A$. For later convenience, the density matrix (\ref{ds2}) can further be diagonalized as $\rho(\tau)=\sum_{i=\pm}\lambda_i|\psi_i(\tau)\ra\la\psi_i(\tau)|$, where
\bea
|\psi_\pm(\tau)\ra
&=&\frac{1}{\sqrt{2|\bm{\rho}|}}\Big(\sqrt{|\bm{\rho}|\mp\rho_3}|0\ra\pm e^{-i\Omega\tau}\sqrt{|\bm{\rho}|\pm \rho_3}|1\ra\Big)\no\\
\lambda_\pm&=&\frac{1}{2}(1\pm|\bm{\rho}|)
\label{ds6}
\eea
with $|\bm{\rho}|=\sqrt{\rho_1^2+\rho_2^2+\rho_3^2}$ is the length of Bloch vector. It was shown \cite{op2} that, in de Sitter space, the detector can reach an equilibrium state $\rho(\infty)$ at later time, which is exactly a thermal state with Gibbons-Hawking temperature $T_H=H/2\pi$.

For instructive reason, we hereafter choose the parameter to be measured as the inverse of Hubble parameter $\beta=2\pi/H$, whose intimate  involvements to cosmological constant $\Lambda$, spacetime curvature $R$, and Gibbons-Hawking temperature $T_H$ are quite obviously.

\section{Quantum estimation for various couplings in de Sitter space }
\label{3}
\subsection{Quantum metrology in de Sitter space}
Our aim is to explore the optimal precision of quantum estimation on the Hubble parameter of de Sitter space. The precision of the estimator is ultimately bounded by QFI, which does not depend on any specific measurement. It can be derived from FI by maximizing it over all possible quantum measurements on the specific parameter $\beta$ introduced before. For a given measurement scheme on the quantum system within state $\rho$, FI relates with a measurement outcome $\xi$ of a positive operator valued measurement (POVM) $\{\hat{E}(\xi)\}$, and takes the form of
\be
\mathcal{F}_C(\beta)=\sum_\xi p(\xi|\beta)\bigg(\frac{\p \ln p(\xi|\beta)}{\p \beta}\bigg)^2\label{ds4}
\ee
where $p(\xi|\beta)$ is the conditional probability of obtaining $\xi$ w.r.t. a chosen POVM and given initial state (\ref{ds2}). From (\ref{ds3}), we observe that the initial states of the detector characterized by $\theta$ and evolving time $\tau$ would play an important role in the metrology process, and eventually determine the ultimate bound on precision. Optimizing (\ref{ds4}) over all the possible quantum measurements of the state (\ref{ds2}), we define the QFI of estimation as $\mathcal{F}_Q(\beta)\equiv\mbox{Max}_{\{\hat{E}(\xi)\}}\mathcal{F}_C(\beta)$, saturated by an \emph{optimal} POVM and can be calculated in terms of the symmetric logarithmic derivative (SLD) operator as $\mathcal{F}_Q(\beta)=\mbox{Tr}[\rho(\beta)L^2_\beta]$, where SLD $L_\beta$ satisfies $\p_\beta\rho=\frac{1}{2}\{\rho,L_\beta\}$. In particular, for a density matrix admitting decomposition (\ref{ds6}), QFI can be further explicitly expressed as \cite{Met26,Met28}
\be
\mathcal{F}_Q(\beta)=\sum_{i=\pm}\frac{(\p_\beta \lambda_i)^2}{\lambda_i}+\sum_{i\neq j=\pm}\frac{2(\lambda_i-\lambda_j)^2}{\lambda_i+\lambda_j}|\la\psi_i|\p_\beta\psi_j\ra|^2\label{ds7}
\ee
where the summations involve sums over all $\lambda_i\neq0$ and $\lambda_i+\lambda_j\neq0$, respectively.

We now apply the detection model to explore the precision bound of quantum estimation in de Sitter space, i.e., to explicitly calculate the QFI relating to Unruh-DeWitt detector with specific dynamics. While any freely falling detector exhibits an \emph{universal} thermal spectrum through interaction with a scalar field in de Sitter space, we further refine detection model by investigate various couplings between scalar field and spacetime curvature, which lead to different Wightman functions. In particular, for a scalar field with Lagrangian $2\mathcal{L}=-g^{\mu\nu}\p_\mu\Phi\p_\nu\Phi-(m_\phi^2+\xi R)\Phi^2$, we consider three kinds of scalar field with distinct couplings, i.e., (i) the conformally coupled massless scalar field with $\xi=\frac{1}{6}$, $m_\phi=0$, (ii) the minimally coupled massless scalar with $\xi=m_\phi=0$, and (iii) the nearly minimally coupled light scalar with $|\xi|\ll1$, $m_\phi\ll H$, respectively. A profound observation reveals \cite{ds16} that the specific choice of couplings may significantly affect the dynamics of entanglement in an expanding spacetime. Therefore, for expanding de Sitter space, it is natural to expect that possible influence on QFI of optimal quantum estimation may occur too.

As a consequence of the amplification of superhorizon modes, the Wightman functions for three kinds of coupling differ strongly at infrared in de Sitter space \cite{ds2,ds3}. Their Fourier transformation can be calculated directly as
\bea
\mathcal{G}_{\mbox{\scriptsize conf}}&=&\frac{\lambda}{2\pi(1-e^{-\beta\lambda})}\no\\
\mathcal{G}_{\mbox{\scriptsize mini}}&=&\frac{\lambda[1+H^2/\lambda^2+2\pi H\delta(\lambda)(H\tau+\mbox{const})]}{2\pi(1-e^{-\beta\lambda})}\no\\
\mathcal{G}_{\mbox{\scriptsize light}}&=&\frac{\lambda[1+H^2/\lambda^2+2\pi H\delta(\lambda)(1/2s+\mbox{const})+\mathcal{O}(s)]}{2\pi(1-e^{-\beta\lambda})}
\label{ds9}
\eea 
with $\beta=2\pi/H$, and the dimensionless parameter is $s=3/2-\nu$ where $\nu=\sqrt{9/4-m^2_\phi/H^2-12\xi}$. For simplicity, we concentrate on the detector without degenerated energy level. Thus the last two expressions of (\ref{ds9}) coincide, and become
\be
\mathcal{G}_{\mbox{\scriptsize mini/light}}=\frac{\lambda[1+H^2/\lambda^2]}{2\pi(1-e^{-\beta\lambda})}\label{ds10}
\ee
Substituting (\ref{ds9}) and (\ref{ds10}) into (\ref{ds11}), we obtain the Kossakowski coefficients as
\bea
A_{\mbox{\scriptsize conf}}&=&\frac{\omega}{4\pi}\frac{e^{\beta\omega}+1}{e^{\beta\omega}-1}~,~A_{\mbox{\scriptsize mini/light}}=\frac{\omega+4\pi^2/\omega\beta^2}{4\pi}\frac{e^{\beta\omega}+1}{e^{\beta\omega}-1}\no\\
B_{\mbox{\scriptsize conf}}&=& \frac{\omega}{4\pi}~,~B_{\mbox{\scriptsize mini/light}}=\frac{\omega+4\pi^2/\omega\beta^2}{4\pi}\no\\
R_{\mbox{\scriptsize conf}}&=&R_{\mbox{\scriptsize mini/light}}=\frac{e^{\beta\omega}-1}{e^{\beta\omega}+1}\label{ds+1}
\eea
which completely determine the final state of Unruh-DeWitt detector (\ref{ds3}). 

\subsection{Fisher information for population measurement.}

The FI can be calculated straightforwardly by (\ref{ds4}). For simplicity, we choose a population measurement on $\beta$, i.e., $\hat{E}_0=|0\ra\la0|$ and $\hat{E}_1=|1\ra\la1|$, which gives outcomes probabilities as $p_i=\mbox{Tr}[\hat{E}_i\rho]$ ($i=0,1$). Then, (\ref{ds4}) becomes 
\be
\mathcal{F}_C(\beta)=\Big[\frac{\p\rho_3}{\p\beta}\Big]^2\Big/(1-\rho_3^2)=\frac{\Big[\Big(\frac{\cos\theta+R}{R} A\tau +1\Big)e^{-A\tau}-1\Big]^2(\p_\beta R)^2}{1-[e^{-A\tau}\cos\theta-R(1-e^{-A\tau})]^2}\label{ds5}
\ee
As mentioned before, (\ref{ds5}) should coincide with QFI calculated from (\ref{ds7}) once the population measurement we chosen is an optimal quantum measurement. 

We now investigate the dependence of FI on the choice of initial state characterizing by $\theta$, the proper time $\tau$ of detector, and the energy level spacing of atom $\omega$, which means FI should be a function as $\mathcal{F}_C(\beta,\theta,\tau,\omega)$, as well as depending on various couplings (\ref{ds9}). Firstly, since $\mathcal{F}_C(\beta)$ (\ref{ds5}) is periodic on $\theta$, one can easily check that the FI of detector would reach its maximum at $\theta=(2n+1)\pi$ ($n\in\mathbb{Z}$), when $(\beta,\tau,\omega)$ and coupling way of scale field are all fixed. On the other hand, as the detector evolves in de Sitter for a long enough time, i.e., $\tau\gg1$ is sufficient larger than the time scale for atomic transition, its final state (\ref{ds3}) would be purely thermalized as a time-independent thermal state \cite{op2} 
\be
\rho(\infty)=\frac{e^{-\beta H_s}}{\mbox{Tr}[e^{-\beta H_s}]}\label{ds12}
\ee
Then, the corresponding asymptotic FI can be given as   
\be
\mathcal{F}_C(\beta)\Big|_{\tau\rar\infty}=\frac{(\p_\beta R)^2}{1-R^2}=\frac{\omega^2}{e^{\beta\omega}+e^{-\beta\omega}+2}\label{ds13}
\ee
which is independent on the choice of initial state, but dependent on the energy spacing $\omega$ and the way of scalar field coupling to curvature. In particular, the asymptotic FI (\ref{ds13}) becomes maximal for certain value $\omega_{max}(\beta)$. Numerically, it is easy to find that for larger Gibbons-Hawking temperature (as $\beta$ degrading), the value of $\omega_{max}$ increases, which is not surprising since a detector in equilibrium with thermal bath at high temperature can be excited to higher energy level.  

\begin{figure}[hbtp]
\includegraphics[width=.8\textwidth]{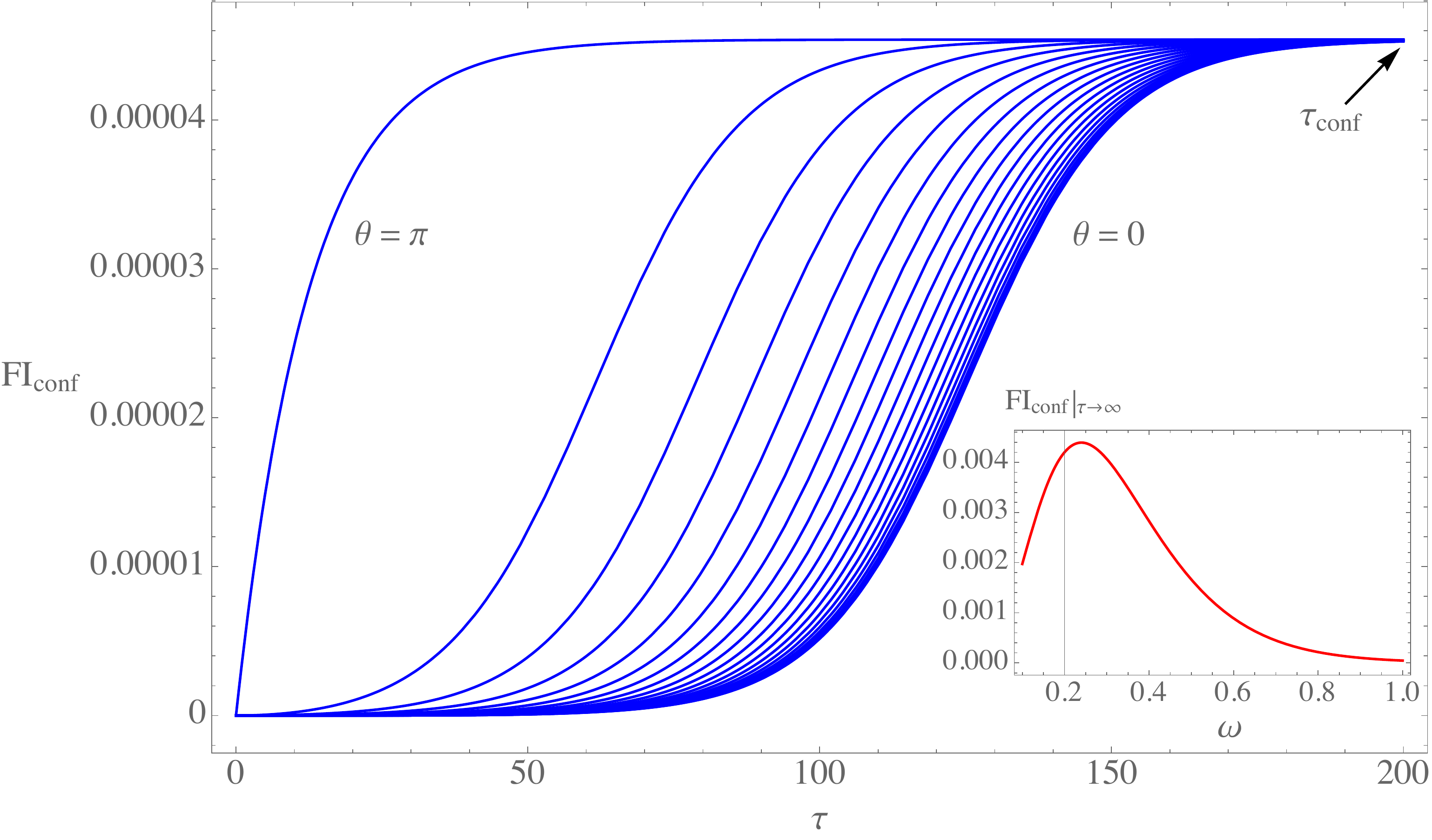}
\caption{For $\beta=10$ and $\omega=1$, FI evolves w.r.t. the proper time $\tau$ of free falling detector interacting with a scalar field conformally coupling to de Sitter. The blue curves indicate the evolution of FI relating to different choice of initial state of detector. The leftmost curve corresponds to $\theta=\pi$, the rightmost one for $\theta=0$, and the difference of $\theta$ for any two neighboring curves is $0.05\pi$. In the inset, the asymptotical FI as a function of $\omega$ is plotted.}
\label{fig1}
\end{figure}

For a freely falling detector with conformally coupled scalar field, we assume $\beta=10$. The corresponding FI evolving w.r.t. proper time $\tau$ is plotted in Fig.\ref{fig1}. For fixed $\omega=1$, while FI with different $\theta$ increasing in distinct rates, they would all asymptotically converge into a fixed value. This indicates that the FI displays a robust \emph{asymptotical maximum} at optimal time $\tau_{\mbox{\scriptsize conf}}\gg1$. In the inset of Fig.\ref{fig1}, we also plot the asymptotical FI as a function of $\omega$, i.e., the energy level spacing of atom.   

\begin{figure}[hbtp]
\includegraphics[width=.8\textwidth]{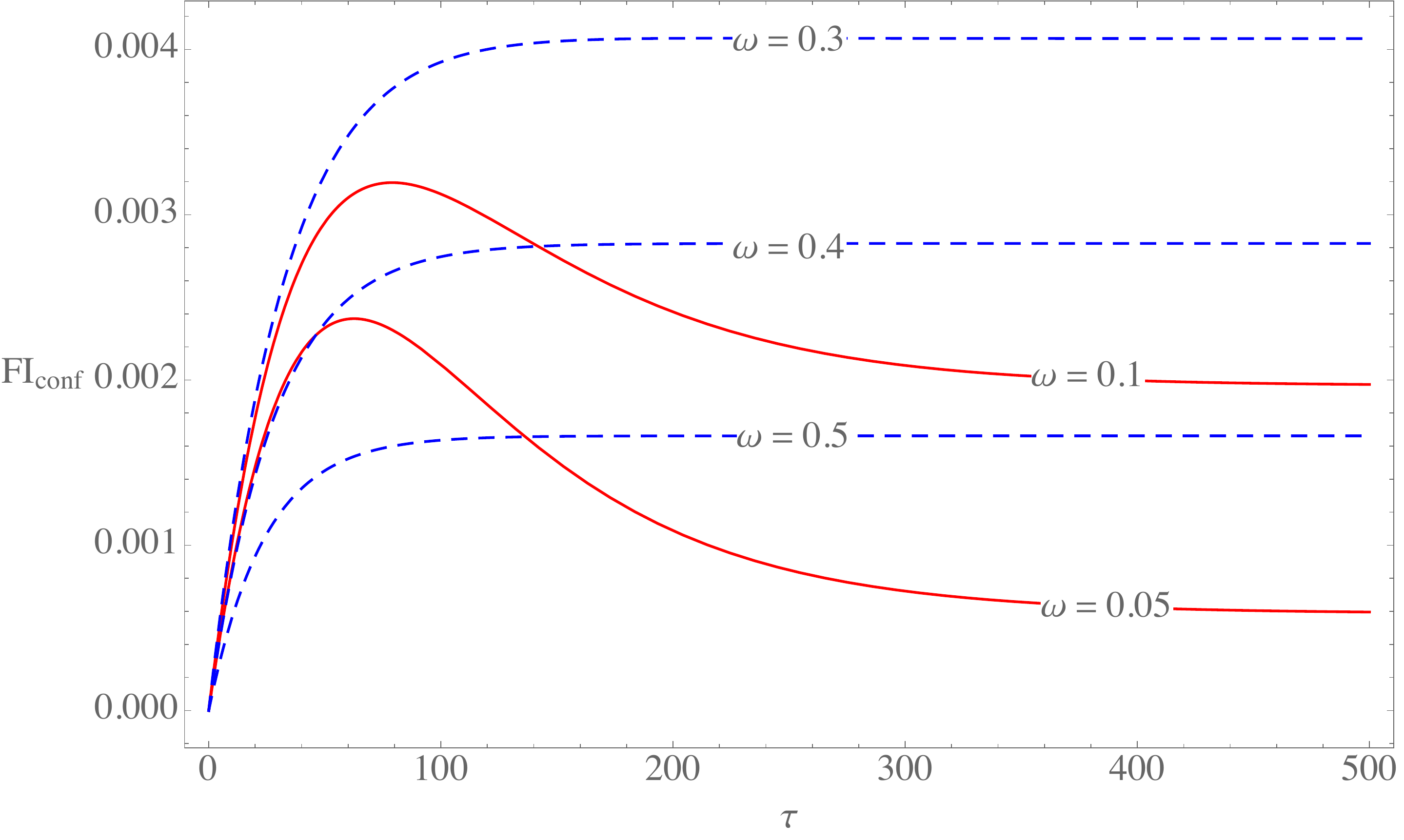}
     \caption{For $\beta=10$ and $\theta=\pi$, FI evolves w.r.t. the proper time $\tau$ of free falling detector interacting with a scalar field conformally coupling to de Sitter. The FI is also the function of the energy level spacing of atom $\omega$. In particular, for certain choice of $\omega$ (e.g., $\omega=0.05$ and $\omega=0.1$), there are some local maxima of FI, which are significantly larger than the corresponding asymptotical maxima. In this case, the global maxima of FI do not monotonously increase w.r.t. the $\omega$.}
\label{fig2}
   \end{figure}

It is interesting to note that, although for any free-falling detector in de Sitter, an asymptotical maximum exists as the detector evolving for enough long time. This does note mean that such asymptotical maxima are also global maxima of FI. As we plot in Fig.\ref{fig2}, for fixed $\beta$ and initial state of detector, there may exist global maximum located at early evolving time. For instance, for $\beta=10$ and $\theta=\pi$, the detector with the energy level spacing $\omega=0.05$ or $\omega=0.1$, exhibits its maximal FI at the time before the optimal time $\tau_{\mbox{\scriptsize conf}}$. Moreover, we should emphasize that the global maxima of FI do \emph{not monotonously} increase w.r.t. the $\omega$, if we specify the coupling of scalar field to de Sitter is conformal. 

\begin{figure}[hbtp]
\includegraphics[width=.8\textwidth]{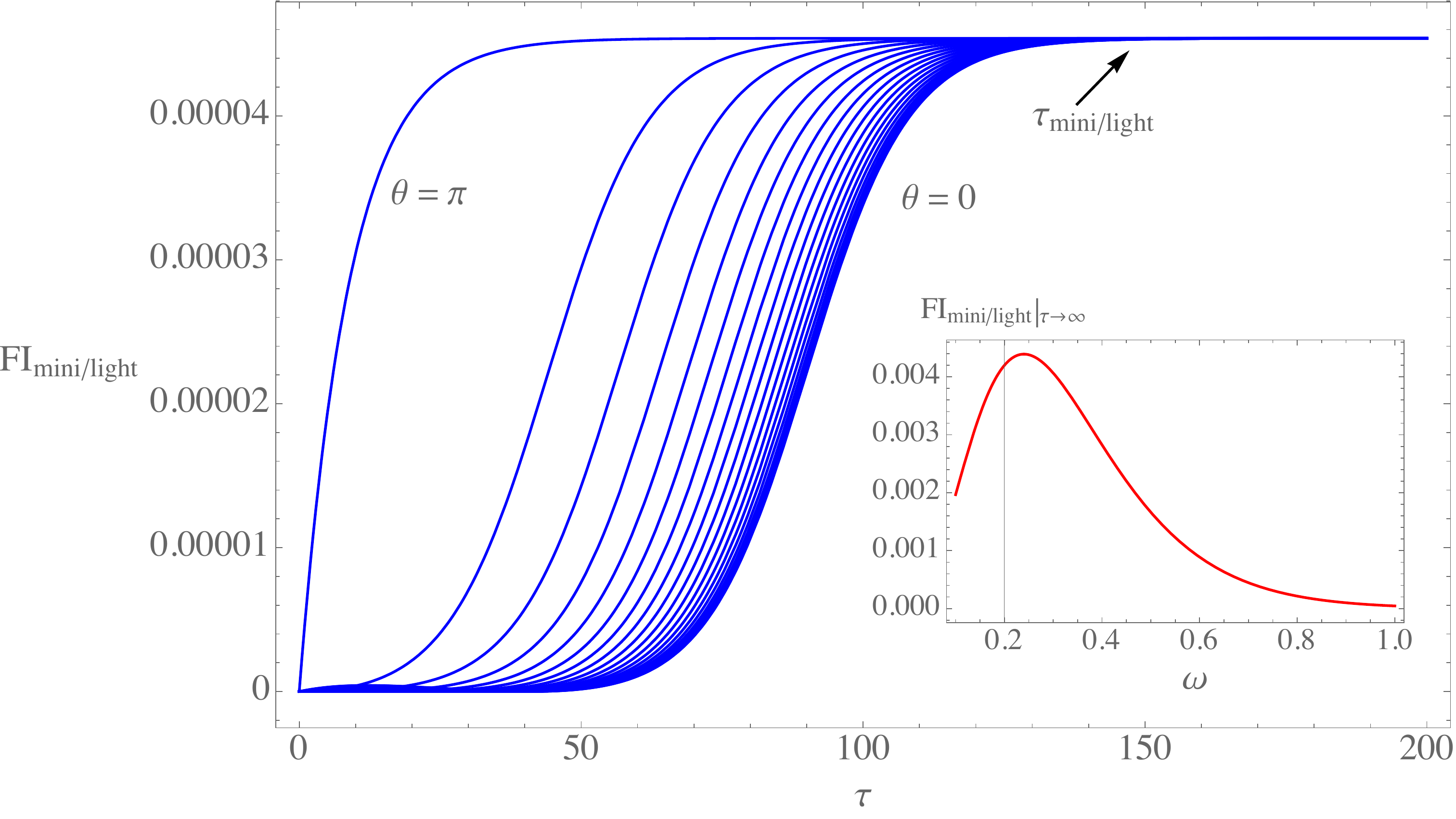}
     \caption{For $\beta=10$ and $\omega=1$, FI evolves w.r.t. the proper time $\tau$ of free falling detector interacting with a scalar field (nearly) minimally coupling to de Sitter. The blue curves indicate the evolution of FI relating to different choice of initial state of detector. The leftmost curve corresponds to $\theta=\pi$, the rightmost one for $\theta=0$, and the difference of $\theta$ of any two neighboring curves is $0.05\pi$. Comparing with Fig.\ref{fig1}, we see a more sharply convergence to the same asymptotical maximum but at an earlier optimal time. In the inset, the asymptotical FI as a function of $\omega$ is plotted.}
\label{fig3}
   \end{figure}

We now turn to the FI of detector interacting with a minimally coupling scalar field or a nearly minimally coupling light scalar field. Substituting (\ref{ds10}) into (\ref{ds5}), we plot the FI of detector evolves w.r.t. the proper time $\tau$ for various initial state choices in Fig.\ref{fig3}. We observe that an asymptotical maximum of FI can be approached at an optimal time $\tau_{\mbox{\scriptsize mini/light}}\gg1$, long enough comparing with the time scale for atomic transition. However, by the choice of scalar field with distinct coupling, we observe that the convergence of FI is much more sharply with minimally coupling/nearly minimally coupling light scalar field. In particular, with same values of $\beta$ and $\omega$, we show that the optimal time $\tau_{\mbox{\scriptsize mini/light}}$, at which the asymptotical maximum appears, is significant smaller than $\tau_{\mbox{\scriptsize conf}}$, meaning a more rapidly stabilizing of FI for the choice of coupling scalar field (\ref{ds10}). This indicates that one may design a quantum estimation with enhanced precision if proper coupling between scalar field and de Sitter space is specified.

\begin{figure}[hbtp]
\includegraphics[width=.8\textwidth]{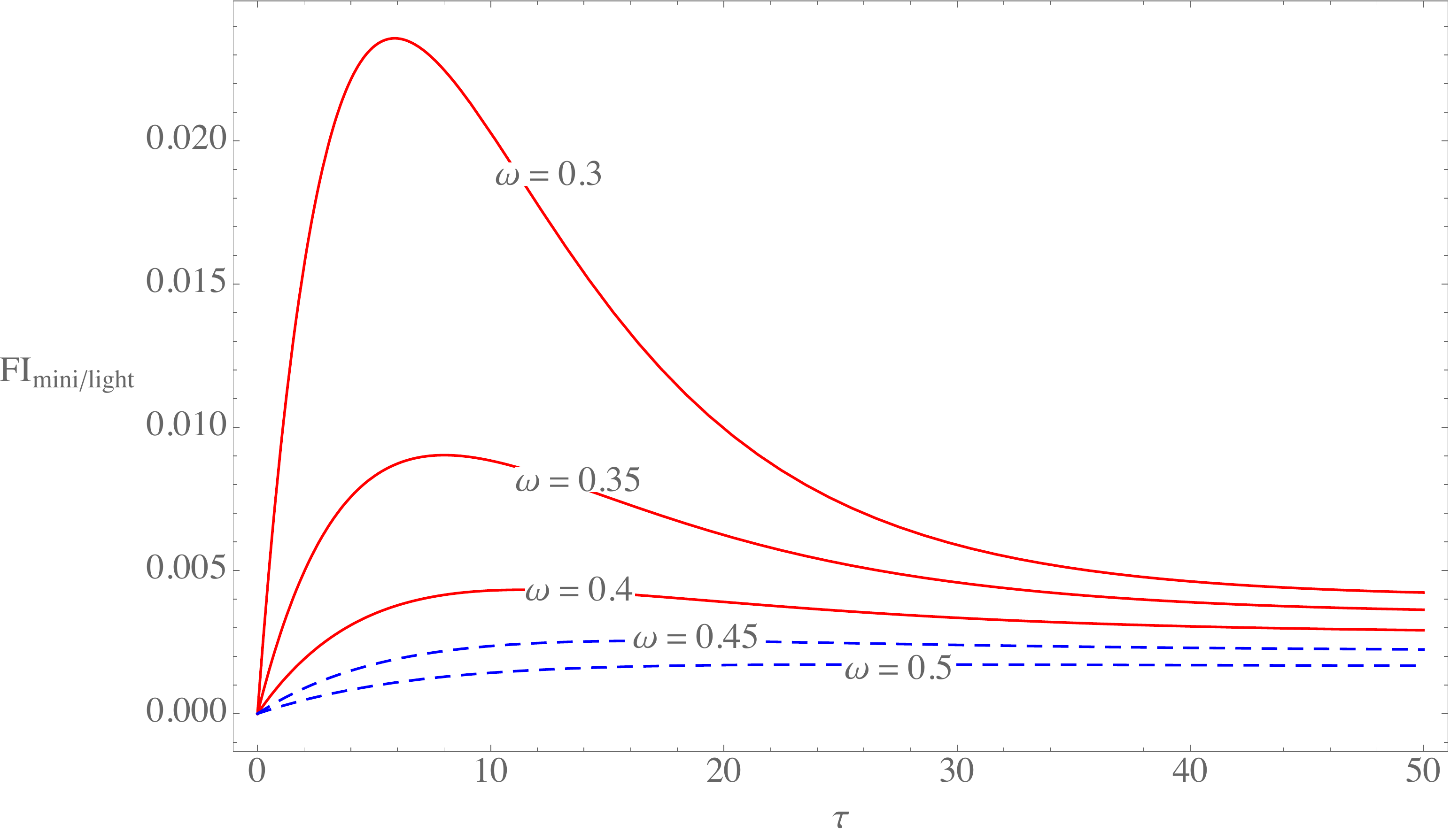}
     \caption{For $\beta=10$ and $\theta=\pi$, FI evolves w.r.t. the proper time $\tau$ of free falling detector interacting with a scalar field (nearly) minimally coupling to de Sitter. The FI is also the function of the energy level spacing of atom $\omega$. In particular, for certain choice of $\omega$ (e.g., $\omega=0.3, 0.35, 0.4$), there are some local maxima of FI, which are significantly larger than the corresponding asymptotical maxima. In opposite to Fig.\ref{fig2}, the global maximum of FI relating to (nearly) minimally coupled scalar field is monotonously decreasing w.r.t. $\omega$. }
\label{fig4}
   \end{figure}

With a choice of scalar field with coupling (\ref{ds10}), for fixed $\beta$ and $\theta$, we plot in Fig.\ref{fig4} the FI evolving w.r.t. the proper time $\tau$, with different values of energy level spacing of detector $\omega$. We observe that for certain values of $\omega$, the FI approaches its maximum earlier than the optimal time $\tau_{\mbox{\scriptsize mini/light}}$, and numerically can be significant larger than the corresponding asymptotical value. Moreover, in contrast to the case with a conformally coupling scalar, such global maximum of FI exhibits a \emph{monotonously degradation} w.r.t. the value of $\omega$, which indicates a lager FI can be achieved for the atomic detector with small energy level spacing.

\begin{figure}[hbtp]
\includegraphics[width=.8\textwidth]{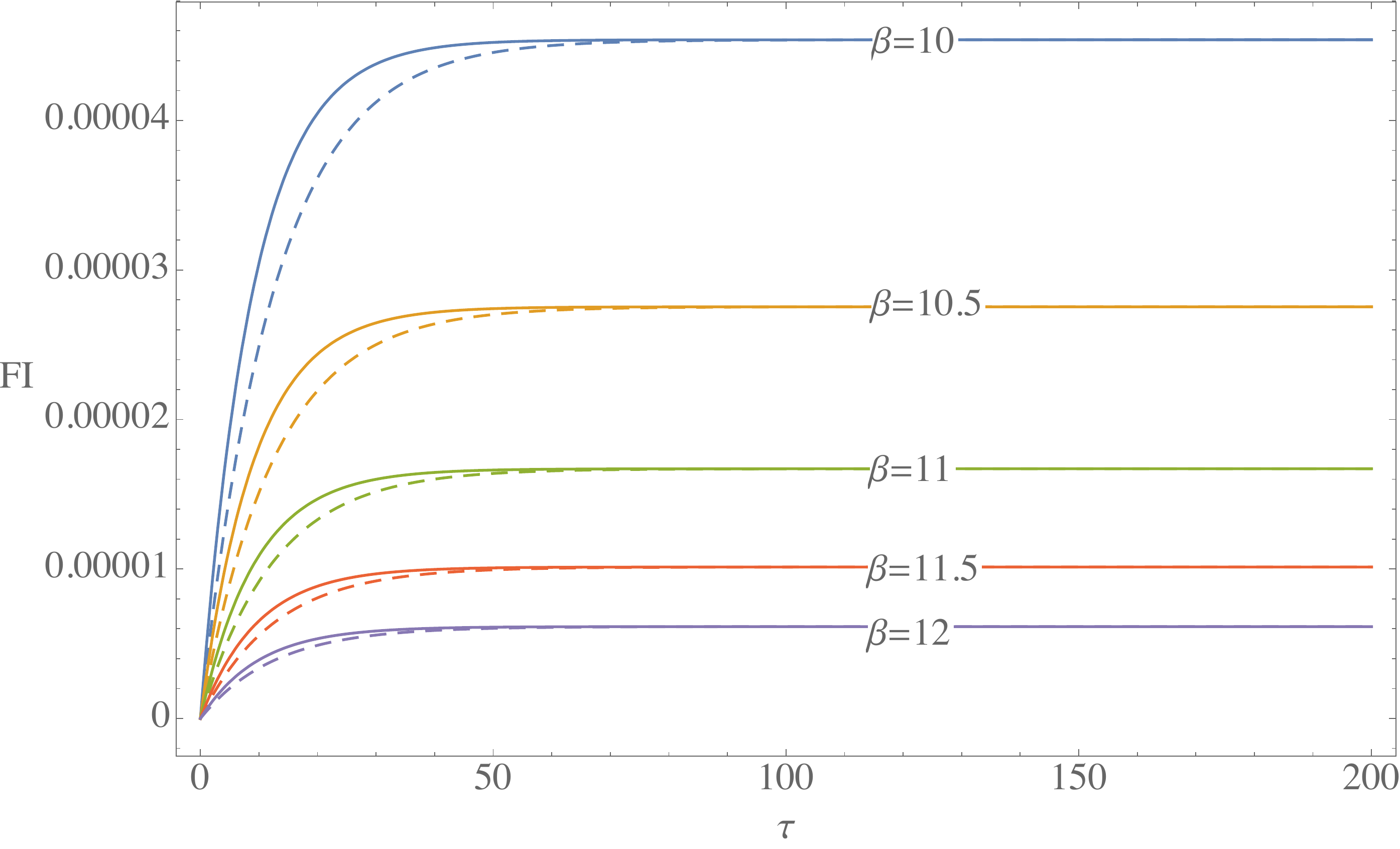}
     \caption{For fixed $\omega=1$ and $\theta=\pi$, FI evolves w.r.t. the proper time $\tau$ of free falling detector in de Sitter space with different Gibbons-Hawking temperature, i.e., $\beta=10,10.5,11,11.5,12$. The solid curves denote the FI for choosing of minimally coupling/nearly minimally coupling light scalar field, while the dashed curves denote the FI relating to conformally coupling scalar field. In both cases, the maximum of FI increase with high temperature. }
\label{fig5}
   \end{figure}

Finally, we plot the FI for different values of Hubble parameter as a function of proper time $\tau$ in Fig.\ref{fig5}. After evolving to the optimal time, the FI approaches its asymptotic value, which degrades as $\beta$ increasing, i.e., it is easer to achieve a given precision estimation in a de Sitter space with larger curvature. 
Moreover, as shown in Fig.\ref{fig5} again, comparing to a choice of conformally coupling scalar field, detector interacting with a minimally coupling/nearly minimally coupling light scalar field (\ref{ds10}) can always achieve a larger FI, indicating a higher precision for estimation. However, as approaching the optimal proper time, FI with different couplings converges into same asymptotic value. To conclude, our result suggests that our ability of high-precision on physical parameters would be enhanced in an expanding spacetime. Moreover, such increment of accuracy can be further improved by fine-tuning of the energy level spacing of detector, as well as by the interaction between the probe and the scalar field with proper coupling to the de Sitter background. 
 
\subsection{Quantum Fisher information}

We now address the evaluation on QFI of the estimation on Hubble parameter. By definition, it can be obtained by optimizing (\ref{ds4}) over all possible quantum measurements. However, since we have already specified population measurement in detection, we alternatively calculate the QFI for the family of states (\ref{ds3}) directly from the more instructive formula (\ref{ds7}), and explore the circumstance under which the QFI can reach its optima. Once the maximal FI can achieve the optimal QFI, we can confirm that the population measurement we chosen is indeed the optimal quantum measurement for estimation process.

Substituting (\ref{ds6}) into (\ref{ds7}), the QFI can be calculated straightforwardly as
\be
\mathcal{F}_Q(\beta)
=\frac{(\p_\beta |\bm{\rho}|)^2}{1- |\bm{\rho}|^2}+\frac{(|\bm{\rho}|\p_\beta\rho_3-\rho_3\p_\beta|\bm{\rho}|)^2}{|\bm{\rho}|^2-\rho_3^2}+\tau^2(|\bm{\rho}|^2-\rho_3^2)(\p_\beta\Omega)^2
\label{ds14}
\ee
where the Lamb contribution in (\ref{ds14}), caused by vacuum fluctuation in curved spacetime, can only be evaluated numerically for de Sitter \cite{ds17,ds18}. Like the FI  (\ref{ds5}), the QFI is a function like $\mathcal{F}_Q(\beta,\theta,\tau,\omega)$, depending on the initial state $\theta$, the proper time $\tau$ of detector, the energy level spacing of atom $\omega$, as well as on various couplings (\ref{ds9}). To access the highest precision of estimation, we have to maximize the value of the QFI over all relevant parameters it depends on. 

\begin{figure}[hbtp]
\includegraphics[width=.8\textwidth]{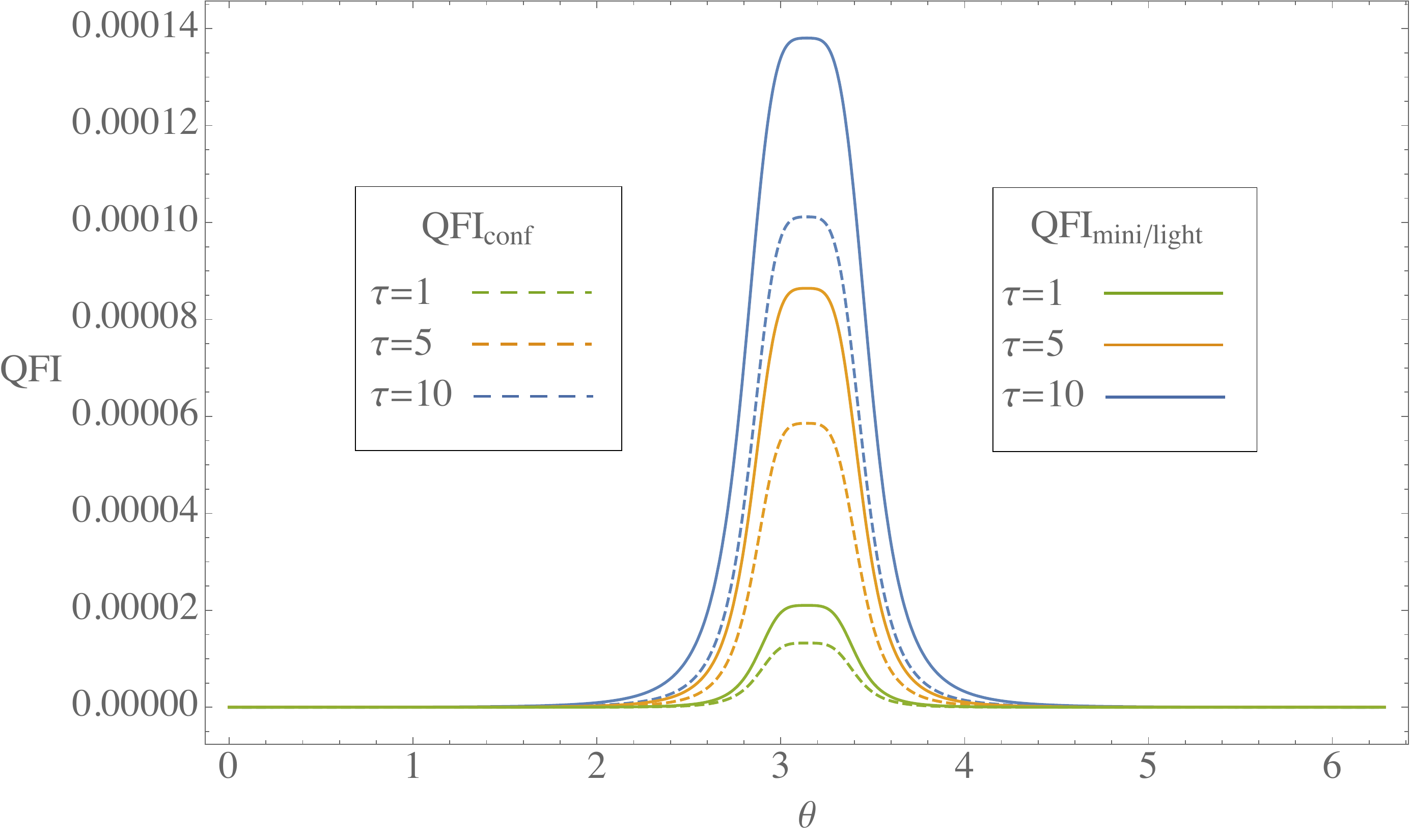}
     \caption{For $\beta=10$ and $\omega=0.8$, QFI evolves w.r.t. the proper time $\tau$ of freely falling detector interacting with a scalar field coupling to de Sitter. The QFI is also dependent on the initial states preparation, labeled by $\theta$. In particular, the maximum of QFI is always located at $\theta=(2n+1)\pi$ ($n\in\mathbb{Z}$).  For fixed $(\beta, \tau, \omega)$, the QFI of estimation relating to minimally coupling/nearly minimally coupling light scalar field can always surpass that relating to conformally coupling scalar field. }
\label{fig6}
   \end{figure}

To find the optimal QFI w.r.t. initial state (\ref{ds4}), in Fig.\ref{fig6}, we assume $\beta=10, \omega=0.8$ and plot the QFI (\ref{ds14}) as a function of $\theta$ for various proper time of free falling detector. We observe that, for two distinct couplings (\ref{ds9}) and (\ref{ds10}) between scalar fields and spacetime, the QFI can always achieve its maxima when $\theta=(2n+1)\pi$ ($n\in\mathbb{Z}$). For same fixed $(\beta, \tau, \omega)$, we find that the optimal QFI with minimally coupling/nearly minimally coupling light scalar fields can always surpass that with conformally coupling scalar field, which means a quantum estimation with a higher efficiency might be designed with the choice of couplings (\ref{ds10}).

Once specifying $\theta=\pi$, (\ref{ds3}) gives $|\bm{\rho}|=\rho_3$. The QFI (\ref{ds14}) can be simplified into 
\be
\mathcal{F}_Q\big|_{\theta=\pi}=\frac{(\p_\beta |\bm{\rho}|)^2}{1- |\bm{\rho}|^2}=\frac{(\p_\beta \rho_3)^2}{1- \rho_3^2}=\mathcal{F}_C\label{ds15}
\ee
which exactly equal to (\ref{ds5}), the FI for population measurement. This just confirms that the population measurement employed for the inertial detector in de Sitter space is indeed an optimal quantum measurement to access the maximal QFI, which leads to the strongest lower bound of the mean-square error in the estimation of Hubble parameter. 
 
Moreover, since both the maximal FI and QFI share the form (\ref{ds15}), the same analysis done for FI (\ref{ds5}) before should also work for the optimal QFI. Conclusively, the dependence of QFI on the parameters $\tau$, $\omega$, $\beta$, as well as the coupling ways of scalar fields to de Sitter background, is just same as described in Fig.\ref{fig2}, Fig.\ref{fig4} and Fig.\ref{fig5}.

\section{Quantum estimation within arbitrary de Sitter-invariant vacua}
\label{4}

So far, we have discussed the FI and QFI in de Sitter space, giving ultimate bound on precision of estimation on Gibbons-Hawking temperature, while the vacuum state of quantum fields are chosen as standard Bunch-Davies vacuum. We now turn to an alternative scenario, in which the Bunch-Davies vacuum are generalized to a family of infinite de Sitter invariant vacua, so-called $\alpha-$vacua. Beside its theoretical interest in high-energy physics (e.g., de/CFT holography), these $\alpha-$vacua are believed to play a significant role in understanding the trans-Planckian physics of early universe.  

\subsection{Dynamics of detector with general $\alpha-$vacua}

For simplicity, we consider the Unruh-DeWitt detector interacting with a massless scalar field conformally coupling to de Sitter space \footnote{Strictly speaking, no de Sitter-invariant vacuum can exist for a massless scalar field \cite{ds5}. However, we can ignore this subtlety by assuming a tiny but nonzero mass \cite{ds20}.}. The Wightman function for the scalar field in $\alpha-$vacua can be expressed in terms of standard Wightman function in Bunch-Davies vacuum as \cite{ds9}
\be
G^+_\alpha(x,y)=\frac{1}{1-e^{2\alpha}}\Big[G^+(x,y)+e^{2\alpha}G^+(y,x)-e^{\alpha}\big(G^+_\alpha(x,y_A)+ G^+(x_A,y)\big)\Big]\label{ds16}
\ee
where the real parameter $\alpha<0$ labels the family of de Sitter invariant vacua, and $x_A$ is the antipodal point of $x$. One can easily find that, as $\alpha\rar-\infty$, $G^+_\alpha(x,y)$ reduces to the Wightman function in Bunch-Davies vacuum $G^+(x,y)$, which uniquely extrapolates to the same short-distance behavior of two-point correlation function in the Minkowski vacuum, as the curvature of de Sitter vanishing. 

To evaluate (\ref{ds8}), we employ the relation 
\be
G^+(x,y_A)=G^+(x_A,y)=G^+(\tau-i\pi)
\ee
and obtain the Fourier transformation of (\ref{ds16}) as
\be
\mathcal{G}^\alpha_{\mbox{\scriptsize conf}}=\frac{\lambda(1+e^{\alpha-\pi\lambda})^2}{2\pi(1-e^{-\beta\lambda})(1-e^{2\alpha})}
\ee
for a conformally coupling scalar field in general $\alpha-$vacua. The related Kossakowski coefficients are 
\bea
A^\alpha_{\mbox{\scriptsize conf}}&=&\frac{\omega\Big[(1+e^{\alpha-\pi\omega})^2+e^{-\beta\omega}(1+e^{\alpha+\pi\omega})^2\Big]}{4\pi(1-e^{2\alpha})(1-e^{-\beta\omega})}\no\\
B^\alpha_{\mbox{\scriptsize conf}}&=&\frac{\omega\Big[(1+e^{\alpha-\pi\omega})^2-e^{-\beta\omega}(1+e^{\alpha+\pi\omega})^2\Big]}{4\pi(1-e^{2\alpha})(1-e^{-\beta\omega})}\no\\
R^\alpha_{\mbox{\scriptsize conf}}&=&\frac{(1+e^{\alpha-\pi\omega})^2-e^{-\beta\omega}(1+e^{\alpha+\pi\omega})^2}{(1+e^{\alpha-\pi\omega})^2+e^{-\beta\omega}(1+e^{\alpha+\pi\omega})^2}\label{ds17}
\eea
which reduce to (\ref{ds+1}) at the limit $\alpha\rar-\infty$. 

Before moving to explore the (Q)FI of detection model, several remarks on $\alpha-$vacua should be addressed. Firstly, any $\alpha$-vacuum $|\alpha\ra$ can be interpreted as a squeezed state over Bunch-Davies vacuum $|0\ra_{BD}$, i.e., $|\alpha\ra=\hat{S}(\alpha)|0\ra_{BD}$ where $\hat{S}(\alpha)$ denotes a squeezing operator in quantum optics \cite{method}. This in general can heavily constrain the measurement uncertainty. Therefore, comparing to the estimation in Bunch-Davies state, such resemblance suggests an outperformance of measurement based on $\alpha-$vacua can be expected in de Sitter space. 

Secondly, $\alpha-$vacua are not thermal in character, following the fact that (\ref{ds16}) does not fulfill the KMS condition unless it reducing to Bunch-Davies vacuum, i.e., $G^+(\tau)=G^+(\tau+i\beta)$. Such deviation from thermality has forced many attempts \cite{ds7,ds8,ds19,ds20} to take $\alpha-$vacua as alternative initial state of inflation. To match the anticipated correction in the primordial power spectrum of order $\sim \mathcal{O}(H/E_P)^2$, the parameter $\alpha$ would exhibit an intimate relation to $E_P$, the energy scale of new physics (e.g., the Planck scale or the stringy scale). For instance, the simplest dependence $e^\alpha\sim(H/E_P)$ constrains the value of $\alpha$ as $\alpha\sim -4$ for Planck scale, $\alpha\sim-2$ for stringy scale.

\subsection{FI and QFI with general $\alpha-$vacua}

\begin{figure}[hbtp]
\includegraphics[width=.8\textwidth]{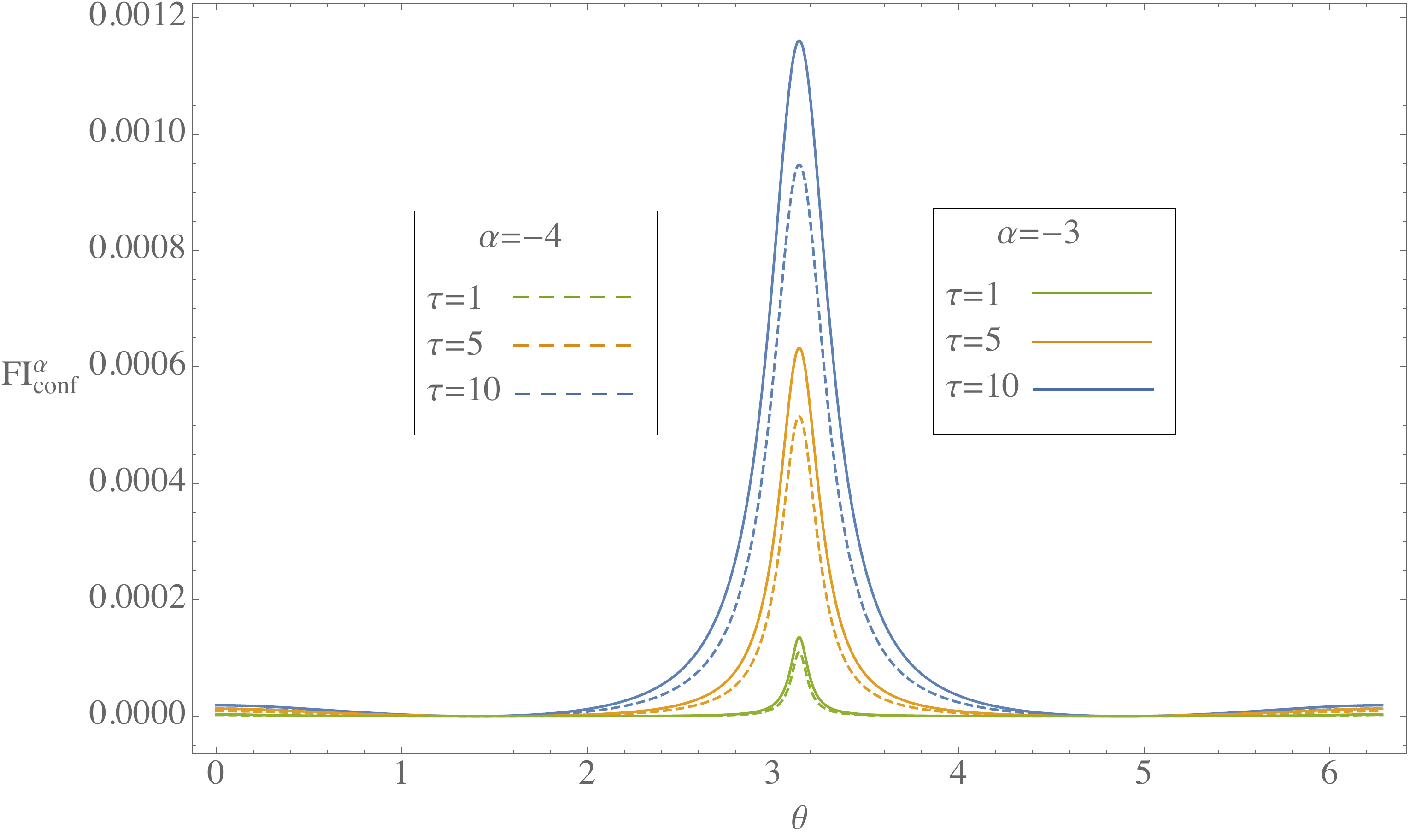}
     \caption{For $\beta=10$ and $\omega=1$, FI of the detector in $\alpha-$vacua is a function of $\theta$, denoting various choices of initial state of detector. The maximum of FI is located at $\theta=(2n+1)\pi$ ($n\in\mathbb{Z}$). For fixed $(\beta, \tau, \omega)$, the maximum of FI located at $\theta=\pi$ increase for larger $\alpha$.}
\label{fig7}
   \end{figure}

We now calculate the FI and QFI for the detection model in $\alpha-$vacua. As the results (\ref{ds5}) and (\ref{ds14}) are universal for relativistic fields in de Sitter space, we can rewrite them in $\alpha-$vacua as
\bea
&&\mathcal{F}^\alpha_C(\beta)=\mathcal{F}^\alpha_Q\big|_{\theta=\pi}(\beta)=\frac{\Big[\Big(\frac{\cos\theta+R^\alpha_{\mbox{\tiny conf}}}{R^\alpha_{\mbox{\tiny conf}}} A^\alpha_{\mbox{\tiny conf}}\tau +1\Big)e^{-A^\alpha_{\mbox{\tiny conf}}\tau}-1\Big]^2(\p_\beta R^\alpha_{\mbox{\tiny conf}})^2}{1-[e^{-A^\alpha_{\mbox{\tiny conf}}\tau}\cos\theta-R^\alpha_{\mbox{\tiny conf}}(1-e^{-A^\alpha_{\mbox{\tiny conf}}\tau})]^2}\no\\
&&\mathcal{F}^\alpha_C(\beta)\Big|_{\tau\rar\infty}=\mathcal{F}^\alpha_Q(\beta)\Big|_{\theta=\pi,\tau\rar\infty}=\frac{(\p_\beta R^\alpha_{\mbox{\tiny conf}})^2}{1-(R^\alpha_{\mbox{\tiny conf}})^2}\label{ds18}
\eea
The FI/QFI are now the functions with further dependence on $\alpha$, i.e., $\mathcal{F}_{C/Q}(\alpha;\beta,\theta,\tau,\omega)$. In the limit of $\alpha\rar-\infty$, the Kossakowski coefficients (\ref{ds17}) reduce to (\ref{ds+1}), therefore the behaviors of $\mathcal{F}_{C/Q}(-\infty;\beta,\theta,\tau,\omega)$ should coincide with those described in Fig.\ref{fig1}-Fig.\ref{fig5}, respectively. In this meaning, it is natural to expect that the previous analysis on the estimation regime in Bunch-Davies vacuum can be inherited for general $\alpha$. For instance, we can numerically confirm that the maxima of FI/QFI in $\alpha-$vacua are located at $\theta=(2n+1)\pi$ ($n\in\mathbb{Z}$). In Fig.\ref{fig7}, we plot FI in $\alpha-$vacua as a function of choice of initial state of detector, for different evolve time $\tau$. Also, numerically, as $\beta$ degrading, we can obtain increasing FI/QFI in $\alpha-$vacua, similar as in Fig.\ref{fig5}. This indicates that the nonvanishing curvature of spacetime can enhance the precision of quantum estimation on Hubble parameter.

\begin{figure}[hbtp]
\includegraphics[width=.8\textwidth]{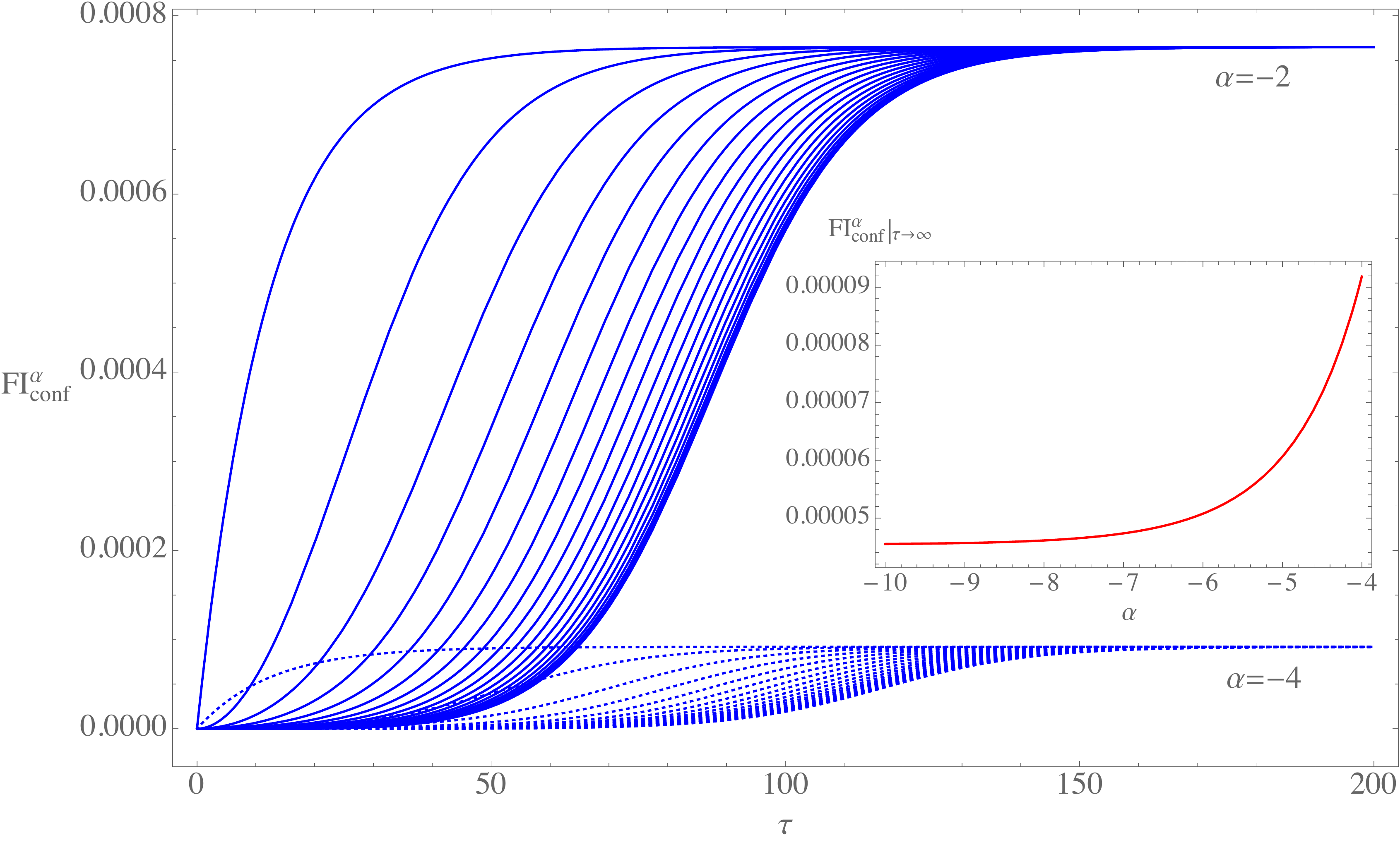}
     \caption{For $\beta=10$ and $\omega=1$, FI evolves w.r.t. the proper time $\tau$ of free falling detector in the $\alpha-$vacua of de Sitter. The solid blue curves indicate the evolution of FI (for $\alpha=-2$) relating to different choice of initial state of detector. The leftmost curve corresponds to $\theta=\pi$, the rightmost one for $\theta=0$, and the difference of $\theta$ of any two neighboring curves is $0.05\pi$. The dotted blue curves indicate the case of FI with $\alpha=-4$. For fixed $\alpha$, the FI converge at certain optimal time to an asymptotic value, which monotonously increase w.r.t $\alpha$ as shown in the inset.
     }
\label{fig8}
   \end{figure}

We are particular interested in the influence of vacua-choice on the value of FI/QFI. As depicted in Fig.\ref{fig7}, for fixed $(\beta_0, \tau_0, \omega_0)$, the maximum of FI/QFI located at $\theta=\pi$ can always be further enhanced by the choice of large $\alpha$, i.e.,
\be
\mathcal{F}_{C/Q}(\alpha;\beta_0,\theta=\pi,\tau_0,\omega_0)>\mathcal{F}_{C/Q}(\alpha';\beta_0,\theta=\pi,\tau_0,\omega_0)\no
\ee
if $|\alpha|<|\alpha'|$ satisfied. In Fig.\ref{fig8}, with $\beta=10$ and $\omega=1$ are fixed, we plot the FI evolving w.r.t. the proper time $\tau$ for different choices of vacuum state, e.g., $\alpha=-2$ and $\alpha=-4$, respectively. We observe a similar convergence of FI for different initial states $\theta$, happening at certain optimal times. Interestingly, the asymptotical value of FI is dependent on the choice of $\alpha$, as it monotonously increases w.r.t $\alpha$ as shown in the inset of Fig.\ref{fig8}. From the relation (\ref{ds18}), we conclude that same enhancement for the maximal QFI (with $\theta=\pi$) should exist for the choice of $\alpha\neq-\infty$. Like we suggested before, such enhancement can be attributed to the squeezing nature of $\alpha-$vacua. Since the squeezing operator $\hat{S}(\alpha)$ can heavily reduce the measurement uncertainty, one can achieve a lower mean-square error in quantum estimation \cite{method}, which means a larger FI/QFI through Cram\'er-Rao inequality. 

We would like to further clarify above result in a context of cosmology. In inflationary paradigm, the non-Bunch-Davies state of quantum fluctuation should be imposed as initial condition of inflation to manifest the possible short-distance cutoff at Planck-scale \cite{ds7,ds8}, that leaves imprint in cosmic microwave background. In this meaning, it is impossible to prepare the initial $\alpha-$vacuum states in order to improve quantum measurements. Nevertheless, since the maxima of QFI (\ref{ds18}) is dependent on the value of $\alpha$, we can conversely regard the optimal QFI as an indicator to distinguish initial states with different $\alpha$.

\section{Conclusions}
\label{6}

In this paper, we investigate the quantum estimation on Hubble parameter of de Sitter space by quantum metrological techniques. By exploring the dynamics a freely falling Unruh-DeWitt detector, which interacts with a scalar field coupling to curvature, we calculate the Fisher information and quantum Fisher information for the estimation of detector, which bound the highest precision one can obtained. In standard Bunch-Davies vacuum of de Sitter, we show that the values of FI/QFI depend on the choice of initial state of probe, the proper time $\tau$ of detector, the energy spacing of atom $\omega$, as well as the way in which the scaler field couples to de Sitter background. In particular, for the initial state prepared with $\theta=\pi$, the FI and QFI of detector can achieve the same maximum for fixed $\tau$ and $\omega$. As $\tau\gg1$, we show a robust asymptotic maximum of FI/QFI exist. Moreover, the maxima of FI/QFI can be further enhanced if proper coupling of scalar field is chosen. To verify this, we compare numerically the FI/QFI for two scenarios in which the detector interacts to scalar field with different couplings to spacetime. We show that with fixed $(\tau,\omega,\theta)$, the estimation in the scenario with minimally/nearly minimally coupling scalar field can always outperform that with conformally coupling scalar field, corresponding to a higher FI/QFI in estimation. Moreover, we find that a further improvement of estimation can be possible if the vacuum state in de Sitter space is chosen as so-called $\alpha-$vacua, which formally can be interpreted as a squeezed state over Bunch-Davies vacuum. Resembling to metrology within quantum optics, we show that the squeezed nature of $\alpha-$vacua can heavily constrain the measurement uncertainty of the detection regime, demonstrated by a higher FI/QFI as $\alpha$ increasing.

The methodology adopted in this paper can be applied to the quantum estimation in other curved spacetime. For instance, one can explore the dynamics of an Unruh-DeWitt detector in the background outside a Schwarzschild black hole, where the detector should be excited by Hawking effect. The estimation on the Hawking temperature would be quite resembling to the estimation of Unruh effect in flat spacetime. Nevertheless, as shown in \cite{ds21+,ds21} that for two Unruh-DeWitt detectors, entanglement can be generated through the evolution of open quantum system, which reduce the entropic uncertainty bound of quantum measurement. Therefore, generalizing present method to Schwarzschild spacetime, it would be interesting to investigate the possible enhanced precision of quantum estimation on the Hawking effect from the entanglement generation in Schwarzschild spacetime.

On the other hand, it would be interesting to extend our model to incorporate more general dynamics of detector. For example, we may include an interaction between detector and scalar field beyond quadratic form in (\ref{model}), which leads to a more complicated dynamics of detector in de Sitter space and the solution of master equation (\ref{ds1}) may only be numerically estimated \cite{nonpert}. In this case, the related FI/QFI might include oscillate behavior w.r.t. proper time due to the involved correlation from quantum vacuum \cite{harve}. Moreover, we would also like to mention that in derivation of the Lindblad equation (\ref{ds1}), a Markov approximation has been invoked, provided that the detector’s temporal resolution is sufficiently large. Nevertheless, such approximation can be eliminated to further include possible back-action of the detector to the field and the spontaneous emission after excitations, i.e., incorporating non-Markovian effects \cite{opentext}. Such effect may be significant in the early-time behavior of FI/QFI, when the detector does not exhibit a thermal behavior as non-Markovian effects are taken into account \cite{nonmarkov}. 

Finally, we have claimed that the enhancement of QFI of quantum estimation in general $\alpha-$vacua could be attributed to the squeezing nature of them. This can be interpreted in a more instructive way. Recall the intimate link of $\alpha$ to the Planck-scale in inflation regime \cite{ds7}, our result indeed implies that the presence of minimal length cut-off in de Sitter could affect the performance of quantum estimation, closely relating to the quantumness (e.g., nonlocality, entanglement, etc) of fundamental scale \cite{ds22}. In fact, for arbitrary spacetime, Planck-scale cut-off would introduce proper deformation in the two-point correlation of underlying field theory \cite{ds23}. Application of our detection model in such scenario would be very interesting, for the exploration on the quantumness of fundamental scale may sheds new light on our understanding of quantum gravity. We will report the related work elsewhere.

\section*{ACKNOWLEDGEMENT}

This work is supported by the National Natural Science Foundation of China (No. 11505133), the Fundamental Research Funds for the Central Universities, Natural Science Basic Research Plan in Shaanxi Province of China (No. 2018JM1049) and the Postdoctoral Science Foundation of China (No. 2016M592769). X.Y.H. acknowledges the support of Innovation and Research Program (No. XJ201710698112). H.F. acknowledges the support of the National Natural Science Foundation of China (No. 91536108, 11774406) and National Key R\&D Program of China (No. 2016YFA0302104, 2016YFA0300600). Y.Z.Z. acknowledges the support of the Australian Research Council Discovery Project (No. 140101492) and the National Natural Science Foundation of China (No. 11775177).\\

\end{document}